\documentclass[sigconf]{acmart}
\copyrightyear{2026}
\acmYear{2026}
\setcopyright{cc}
\setcctype{by}
\acmConference[CCS '26]{Proceedings of the 2026 ACM SIGSAC Conference on Computer and Communications Security}{November 15--19, 2026}{The Hague, Netherlands}
\acmBooktitle{Proceedings of the 2026 ACM SIGSAC Conference on Computer and Communications Security (CCS '26), November 15--19, 2026, The Hague, Netherlands}
\acmDOI{10.1145/3830454.3832688}
\acmISBN{979-8-4007-2871-6/2026/11}

\author{Rizky~Ramadhana Putra}
\affiliation{%
  \institution{Virginia Tech}
  \city{Blacksburg}
  \state{VA}
  \country{USA}}
\email{rizky@vt.edu}

\author{Osama Bajaber}
\affiliation{%
  \institution{King Abdulaziz University}
  \city{Jeddah}
  \country{Saudi Arabia}}
\email{obajaber@kau.edu.sa}

\author{Saimon~Amanuel~Tsegai}
\affiliation{%
  \institution{Virginia Tech}
  \city{Blacksburg}
  \state{VA}
  \country{USA}}
\email{saimon.tsegai@vt.edu}

\author{Teryl Taylor}
\affiliation{%
  \institution{IBM Research}
  \city{Yorktown Heights}
  \state{NY}
  \country{USA}}
\email{terylt@ibm.com}

\author{Frederico Araujo}
\affiliation{%
  \institution{IBM Research}
  \city{Yorktown Heights}
  \state{NY}
  \country{USA}}
\email{frederico.araujo@ibm.com}

\author{Yuede Ji}
\affiliation{%
  \institution{UT Arlington}
  \city{Arlington}
  \state{TX}
  \country{USA}}
\email{yuede.ji@uta.edu}

\author{Peng Gao}
\affiliation{%
  \institution{Virginia Tech}
  \city{Blacksburg}
  \state{VA}
  \country{USA}}
\email{penggao@vt.edu}

\usepackage{tikz}
\usepackage{amsmath}

\usepackage{filecontents}
\usepackage{tablefootnote}

\usepackage{microtype}
\usepackage{algorithm}
\usepackage[spaceRequire=false]{algpseudocodex}
\usepackage[utf8]{inputenc} 
\usepackage{listings}
\usepackage{xspace}
\usepackage{subcaption}
\usepackage{soul}
\usepackage{enumitem}
\usepackage{booktabs}
\usepackage{multirow}
\usepackage{tcolorbox}
\usepackage[subtle]{savetrees}
\usepackage{makecell}
\usepackage{balance}

\renewcommand{\shortauthors}{Rizky Ramadhana Putra et al.}

\usepackage[labelfont=bf,skip=5pt,figurename=Fig.]{caption}
\def\circled#1{%
  \hspace{0.2em}
  #1%
  \pdfliteral{
    q .5 w
    10 0 0 10 -2.5 3.5 cm .05 w .5 0 m
    .5 .276 .276 .5 0 .5 c -.276 .5 -.5 .276 -.5 0 c
    -.5 -.276 -.276 -.5 0 -.5 c .276 -.5 .5 -.276 .5 0 c h
    S Q
  }%
  \hspace{0.2em}
}

\usepackage[capitalize,nameinlink]{cleveref}
\hypersetup{%
	bookmarksnumbered, bookmarksopen=true, bookmarksopenlevel=1,%
}
\crefname{figure}{Fig.}{Figs.}
\crefname{listing}{Listing}{Listings}
\crefname{section}{\S}{Sections}
\crefname{table}{Table}{Tables}
\crefname{algorithm}{Algorithm}{Algorithms}
\crefname{line}{Line}{Lines}
\crefname{equation}{Eq.}{Eqs.}
\crefname{rq}{RQ}{RQs}
\crefname{appendix}{Appendix}{Appendices}

\newcommand{\myparagraph}[1]{\smallskip\noindent\textbf{#1.}}

\newcommand{\tool}{\textsc{eMicro}\xspace}
\newcommand{\toolnospace}{\textsc{eMicro}}

\newif\ifshowcomment
\showcommenttrue

\ifshowcomment
    \newcommand{\pgao}[1] {{\footnotesize\color{red}[Peng: #1]}}
    \newcommand{\osama}[1] {{\footnotesize\color{orange}[Osama: #1]}}
    \newcommand{\saimon}[1] {{\footnotesize\color{cyan}[Saimon: #1]}}
    \newcommand{\rizky}[1]{{\footnotesize\color{purple}[Rizky: #1]}}
    \newcommand{\teryl}[1]{{\footnotesize\color{green}[Teryl: #1]}}
    \newcommand{\fred}[1]{{\footnotesize\color{magenta}[Fred: #1]}}
\else
    \newcommand{\pgao}[1]{}
    \newcommand{\osama}[1]{}
    \newcommand{\saimon}[1]{}
    \newcommand{\rizky}[1]{}
    \newcommand{\teryl}[1]{}
    \newcommand{\fred}[1]{}
    
\fi

\newcommand{\eat}[1]{}

\begin{document}

\title{\tool: Real-Time Multi-Hop Access Control for Microservices with eBPF}

\begin{abstract}
  
Modern cloud applications often comprise thousands of microservices whose interactions form complex request paths. 
Traditional inter-service access control restricts individual service-to-service requests, but fails to prevent \emph{multi-hop attacks}, where each hop appears legitimate yet the overall path violates security intent. This gap leaves systems exposed to unauthorized access and data exfiltration.
In this paper, we present \tool, a path-aware defense system for microservices that prevents such attacks while remaining efficient and deployable.
\tool enforces real-time multi-hop access control through three key techniques:
(1)~history-based access control extended to capture service invocation sequences;
(2)~security policies encoded as efficient deterministic finite automaton (DFA), 
supporting constant-time lookups and compact label propagation;
(3)~eBPF-based in-kernel request tracing for transparent, low-overhead enforcement without code changes.
Evaluations on DeathStarBench and production cloud traces from Uber, Alibaba, and ByteDance, covering 12 million request workflows and thousands of services, demonstrate the scalability of \tool.
\tool performs policy checks in 1 $\mu$s, stores 50 million policies in only 100 MB, and reduces propagation overhead by 90\% with negligible runtime impact.
These results show that \tool delivers scalable and efficient protection against multi-hop attacks, making it practical for deployment in large-scale microservice environments.

\end{abstract}



\maketitle


\section{Introduction}

Microservices decompose modern applications into loosely coupled services that communicate over the network. 
At scale, such systems may consist of thousands of services, producing millions of interdependent requests and intricate request sequences. 
This complexity enlarges the attack surface by introducing opportunities for \emph{multi-hop attacks}~\cite{cloudcover}, in which adversaries exploit legitimate inter-service dependencies to construct request chains that are syntactically valid but semantically violate security intentions. 
This \emph{generic threat} manifests in diverse forms across modern microservice deployments.
For example, Datadog reported cryptojacking campaigns in Docker Swarm and Kubernetes \cite{datadog}, where attackers first compromised unauthenticated Internet-facing services and then abused service dependencies to hijack resources for cryptocurrency mining.

Existing defenses largely decouple inter-service authorization from intra-service enforcement. 
Intra-service defenses (e.g., system call specialization~\cite{upolicycraft,confine}) constrain the execution of individual services but ignore requests exchanged across services, and thus cannot associate execution spanning multiple services. 
Inter-service defenses (e.g., network policies and inter-service access control~\cite{bastion,autoarmor,jarvis,eztrust,log2policy,cilium,calico,istio}) regulate communication between pairs of services while overlooking each service's internal execution, and are unable to capture the causal dependencies needed to connect a chain of service interactions. 
Therefore, neither class can effectively prevent multi-hop attacks.
Provenance-based defenses for microservices~\cite{clarion,winnower,camflow} can reconstruct complete request sequences, but their use is largely limited to post-attack forensic analysis rather than real-time prevention. 
Information-flow-based defenses~\cite{trapeze,fabric} extend visibility beyond adjacent hops, but do not capture the precise ordering of request sequences.

Early approaches toward real-time multi-hop enforcement~\cite{cloudcover,william,sidecar,expressive} suffer from scalability limitations. 
They rely on \emph{centralized servers} that introduce bottlenecks and single points of failure, incur millisecond-level delays, and do not support concurrent requests efficiently. 
Many also depend on specific frameworks or libraries, resulting in ad hoc solutions.
Consequently, the problem of preventing multi-hop attacks in real time, while ensuring scalability, efficiency, and deployability, has not been fully addressed.

\myparagraph{Goals and challenges} 
To mitigate multi-hop attacks in microservices, we aim to design a \emph{real-time defense} that enforces policies over \emph{entire} request sequences rather than individual hops. 
Achieving this goal requires overcoming three key challenges.

\emph{\textbf{First}, 
Preventing multi-hop attacks requires validating the entire invocation history up to the originating service}, since intermediate hops often encode critical security intent. 
However, naively enumerating all permissible request sequences quickly leads to policy explosion, as real deployments exhibit a combinatorial number of workflows, making policies brittle and difficult to maintain. The challenge is to support policies that can concisely constrain request paths, such as requiring, forbidding, or matching specific intermediate services, without 
sacrificing administrative usability.

\emph{\textbf{Second}, enforcing multi-hop policies must be real-time and scalable.}
This requires the enforcement mechanism to track the sequence of preceding requests while remaining lightweight enough for real-time operation.
Naively attaching the complete preceding request sequences to each request and sequentially checking them against all policies incurs $\mathcal{O}(n)$ time complexity in the number of policies, and does not scale when thousands of access control decisions  performed per second.
Production cloud traces~\cite{uber_ds, ali_ds, ali_ds2, bd_ds, meta} report deployments with up to 40{,}000 services and request paths of length up to 80, yielding large per-request context.

Prior multi-hop microservice access control approaches~\cite{cloudcover,kalium,william} rely on a centralized remote controller to reconstruct and evaluate request histories across the cluster. This adds a network round-trip on every hop and creates bottlenecks and a single point of failure.
A natural attempt is to replicate the controller via a consensus protocol like Raft~\cite{raft}, but this does not remove the bottleneck. Raft makes the controller highly available, yet the cost comes from per-request evaluation over a causal chain whose events are produced at every upstream service. However the controller is replicated, requests tracing and policy enforcement must still reach the decision point, either through a per-hop round trip or through per-request Raft consensus, beyond the update rate Raft is designed for.

Modern orchestration platforms such as Kubernetes further complicate this challenge: services are dynamically scheduled, replicated, and migrated across physical machines, with underlying placement hidden behind service discovery, overlay networks, service meshes, and sidecars, so a service rarely knows which physical host its peer runs on. 
A user-space observer is often insufficient, since namespace isolation limits visibility and frequent user/kernel context switches add substantial overhead. 

\emph{\textbf{Third}, microservices are implemented using diverse languages, libraries, frameworks, and parallelism paradigms.}
They communicate over heterogeneous protocols (e.g., HTTP, RPC) and middleware stacks (e.g., Thrift, gRPC, Avro).
Propagating request-sequence context across workflows requires instrumenting applications or the communication libraries and frameworks they depend on, incurring substantial engineering overhead.
This ecosystem is also large, rapidly evolving, and highly heterogeneous.
For example, Meta reports production use of 16 programming languages~\cite{meta}, while Jaeger and OpenTelemetry \cite{jaeger} maintain instrumentation across 11 languages, each with multiple dialects (e.g., Meta’s Hack).
An effective solution therefore must propagate hop information and enforce policies 
in a \emph{communication-stack-agnostic manner}, without requiring much application code modification.

\myparagraph{Contributions}
We present \textbf{\tool, a decentralized multi-hop access control system for microservices} that secures entire request workflows by enforcing end-to-end policies.
\tool prevents multi-hop attacks in real time using a path- and order-aware policy language that is compiled into a compact DFA, distributed across services, enforced at line rate with minimal overhead, and requires no manual application or library instrumentation while remaining compatible with common cloud networking platforms.
\tool advances microservices security beyond inter-service single-hop access control \cite{cilium, calico, istio, bastion, eztrust} and information-flow-based defenses that do not enforce request order \cite{trapeze, fabric, p4control}. 
Compared with prior centralized multi-hop access control systems \cite{cloudcover, expressive, sidecar, william, kalium}, \toolnospace's primary contribution lies in scalable, instrumentation-free, decentralized enforcement rather than new policy semantics.

Our contributions are threefold:

\begin{enumerate}[label={\textbf{(\arabic*)}}, leftmargin=*, itemsep=6pt, topsep=6pt]
        
    \item \emph{Expressive multi-hop policy language.}
    \tool introduces a policy language that models benign request workflows and prevents multi-hop attacks by reasoning about complete request paths rather than only adjacent service hops.
    Instead of enumerating all valid workflows, the language provides path-level predicates and logical operators that compactly capture security intent over intermediate services.
    By going beyond simple sequence matching, it avoids policy explosion while enabling rich security semantics.
    As a result, the language remains practical and easy for administrators to author and maintain.

    \item \emph{Compact DFA-based representation for multi-hop policies.}
    \tool avoids propagating full request histories and sequentially checking policies, which does not scale and precludes real-time enforcement.
    Instead, \tool compiles multi-hop policies describing valid request workflows into a single deterministic finite automaton (DFA) with \emph{constant-time} complexity policy check, where each state represents a valid request sequence.
    \emph{Requests propagate only the corresponding DFA state label}, rather than complete histories, yielding significantly smaller labels both theoretically and in practice.
    To support large policy sets, our compiler employs a data-parallel algorithm that merges thousands of policy DFAs efficiently using a tree-structured parallel reduction.
    
    \item \emph{Application-agnostic label propagation and policy enforcement via eBPF.}
    \tool propagates DFA state labels and enforces multi-hop policies \emph{transparently} without modifying application code.
    Leveraging eBPF technology~\cite{ebpf}, \tool operates at the kernel level, enabling low-overhead label propagation across network packets, processes, threads, and coroutines, while avoiding user-kernel context switches.
    Policies are replicated across nodes using consensus protocols (Raft~\cite{raft}), while enforcement remains local at each service and decentralized.

\end{enumerate}

\begin{figure*}[t]
    \centering
    \begin{subfigure}[]{0.33\linewidth}
        \includegraphics[scale=1.15]{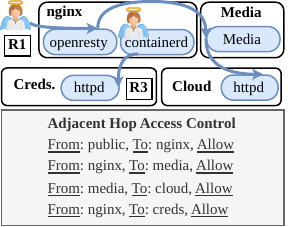}
        \caption{}\label{fig:correct_request_workflow}
    \end{subfigure}
    \hfill
    \begin{subfigure}[]{0.33\linewidth}
        \includegraphics[scale=1.15]{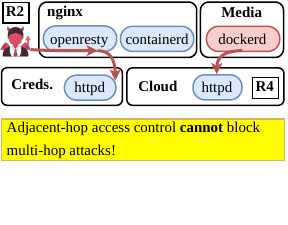}
        \caption{}\label{fig:adjacent_hop_access_control}
    \end{subfigure}
    \hfill
    \begin{subfigure}[]{0.33\linewidth}
        \includegraphics[scale=1.15]{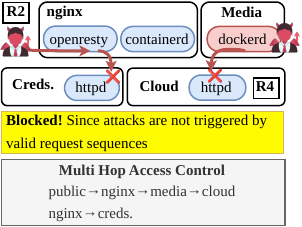}
        \caption{}\label{fig:multihop_access_control}
    \end{subfigure}
    
    \caption{Two valid request workflows (\fbox{R1}, \fbox{R3}). A developer uses a malicious Docker image that performs network enumeration (\fbox{R2}). An attacker exploits a server-side request forgery (SSRF) vulnerability in \texttt{nginx} to send requests to \texttt{creds} (\fbox{R4}). Both \fbox{R3} and \fbox{R4} send requests from \texttt{nginx} to \texttt{creds}, and both \fbox{R1} and \fbox{R2} send requests from \texttt{media} to \texttt{cloud}. Adjacent-hop access control treats these cases identically, enabling multi-hop attacks due to missing end-to-end context.}
    \label{fig:example}
\end{figure*}


\myparagraph{Evaluation}
We conducted extensive evaluations on a 9-VM deployment and evaluated \tool using large-scale production traces from Alibaba~\cite{ali_ds2,ali_ds}, Uber~\cite{uber_ds}, and ByteDance~\cite{bd_ds}.
We also deployed \tool on the DeathStarBench~\cite{deathstarbench} testbed 
running 27 services with 10 replicas each and processing 12 million requests over 12 hours under realistic network conditions (latency, jitter, packet loss).
The testbed reflects real-world microservice deployments, spanning six programming languages (Java, Node.js, Python, C/C++, JavaScript, Go), open-source services (NGINX, Memcached, MongoDB, Cylon, Xapian, MySQL, RabbitMQ), common protocols (HTTP, RPC), frameworks (Apache Thrift, gRPC), and custom user-defined applications.
In addition to benign workloads, we evaluated representative multi-hop attack scenarios, including open-redirection attacks, remote code execution (RCE)-based workflow bending, and malicious container-based network enumeration.
We also compared \tool with prior centralized multi-hop enforcement systems (CloudCover~\cite{cloudcover}) and information-flow-based defenses (Kalium~\cite{kalium} and Trapeze~\cite{trapeze}).

Our evaluation results demonstrate that: 
(1) \tool can effectively block all malicious flows in real-time while incurring \textbf{less than 10~$\mu$s per-request overhead}.
(2) \toolnospace's DFA-based compilation reduces the space required to propagate request sequences by 90\% on average, requiring only $\sim$100~MB to store 50~million policies with constant-time lookups under 1~$\mu$s.
(3) \toolnospace's fully distributed enforcement achieves \textbf{35\% shorter completion times} compared to CloudCover~\cite{cloudcover}, and performs policy checks up to 10$\times$ faster than Kalium~\cite{kalium} and Trapeze~\cite{trapeze}, incurring negligible overhead to legitimate traffic.
(4) To further demonstrate \toolnospace's robustness under failures, we integrate \tool with the Raft consensus protocol~\cite{raft} for policy distribution. 
Our results show that policy updates reach quorum in under 4~seconds, and failed nodes recover missing updates within an additional 4~seconds, ensuring fast failover recovery and consistent policy enforcement across nodes.
These results demonstrate that \tool effectively prevents multi-hop attacks while remaining efficiently deployable, non-intrusive, and minimally disruptive to benign workloads.


\section{Background and Related Work} \label{sec:background}

{Microservices} is an architectural style in which applications are decomposed into small, loosely coupled services that interact over the network using protocols such as HTTP or RPC. Each service typically executes inside a container managed by a runtime such as Docker \cite{docker} or containerd \cite{containerd}.
While containers share the same underlying kernel, they maintain isolation in networking, file system, and processes through namespaces. Deployments at scale are orchestrated by platforms such as Docker Swarm \cite{docker} and Kubernetes \cite{kubernetes}, which establish virtual networking between services, commonly implemented through iptables or container network interface (CNI) plugins such as Cilium \cite{cilium} and Calico \cite{calico}.

\subsection{Motivating Example}
\label{sec:motivation_example}

To illustrate the limitations in microservice access control mechanisms against multi-hop attacks, consider a web-based photo-sharing application composed of four services: an \verb|nginx| frontend, an image processing service \verb|media|, a \verb|cloud| storage, and a credential management service \verb|creds|. \cref{fig:correct_request_workflow} shows the two legitimate workflows supported by the application.
In the first workflow, a user loads a webpage containing a picture. The browser sends the request to \verb|nginx|, which forwards it to the \verb|media|. The \verb|media| retrieves the image from \verb|cloud|, applies transformations such as resizing and compression, and returns the processed image to the user via \verb|nginx|. In the second workflow, the \verb|nginx| frontend server periodically retrieves short-lived tokens from \verb|creds|, which are used to securely access \verb|media| and \verb|cloud| without embedding long-term secrets.

Now suppose the \verb|nginx| contains a server-side request forgery (SSRF) vulnerability. An attacker could exploit this flaw to redirect requests so that \verb|nginx| forwards them to \verb|creds| instead of \verb|cloud|. This enables the attacker to extract authentication tokens and subsequently access other services. Conventional microservice access control considers only adjacent hops and cannot distinguish between requests \emph{initiated directly} by the \verb|nginx| and those \emph{triggered indirectly via} \verb|nginx|, as illustrated in \cref{fig:adjacent_hop_access_control}.
Similarly, a compromised \verb|media| container that performs network enumeration and sends requests to \verb|cloud| cannot be prevented.

Preventing such attacks requires reasoning about entire request sequences, not just individual hops. For example,  \verb|nginx| should access \verb|creds| only when doing so directly, not when acting on behalf of a user request. 
Similarly, \verb|media| may access \verb|cloud| only if initiated by \verb|[public,nginx]| request sequences.
This approach, depicted in \cref{fig:multihop_access_control}, follows a history-based access control paradigm \cite{fournet2003access,10.1145/288090.288102,1301314} in which actions are authorized based on their causal history rather than in isolation. In microservices, the relevant history corresponds to the sequence of requests recursively triggered to fulfill a workflow. Authorization decisions should therefore derive from the complete preceding request sequence.

\subsection{Multi-Hop Attacks}

Multi-hop attacks are categorized into two broad classes: path-suffix attacks, which exploit fragments of benign workflows, and workflow-bending attacks, which construct entirely new workflows that are not aligned with legitimate ones~\cite{cloudcover}.

\textbf{(i) Path-suffix attack.}
Consider again the photo-sharing application (\cref{fig:example}). Two valid workflows exist: \verb|[public,nginx,media,| \verb|cloud]| and \texttt{[nginx,} \texttt{creds]}. Suppose an attacker compromises \verb|media| and directly invokes \verb|cloud|, creating the path \texttt{[media,} \texttt{cloud]}. Although this sequence resembles part of a benign workflow, it is invalid because it bypasses the required initiation from \verb|public| and \verb|nginx|. Path-suffix attacks are feasible in practice through software supply-chain vulnerabilities: malicious or trojanized container images may embed scripts that enumerate and contact other services. Prior work shows that more than 80\% of images on Docker Hub contain at least one high-severity vulnerability \cite{dockerhub}, making this a significant attack vector.

\textbf{(ii) Request-workflow-bending attack.}
Workflow-bending attacks deviate entirely from valid workflows \cite{cloudcover}. In some cases, attackers exploit a benign service as a confused deputy. For example, an open-redirection vulnerability in \verb|nginx| \cite{openredirection} could be abused to bend the workflow from \texttt{[public,} \texttt{nginx,} \texttt{media,} \texttt{cloud]} to \texttt{[public,} \texttt{nginx,} \texttt{creds]} (\cref{fig:adjacent_hop_access_control}). In other cases, attackers compromise a service, allowing them to inject malicious processes and generate arbitrary inter-service requests\cite{mitre}.

\subsection{Microservices Access Control}
\label{sec:hbac}

Below, we review prior work on microservice access control and contrast it with our approach.

\begin{itemize}[align=left, labelsep=0pt,labelwidth=\parindent,leftmargin=14pt,itemsep=6pt, topsep=6pt]
    \item \textit{Adjacent-hop enforcement.} Approaches such as iptables, Cilium \cite{cilium}, Calico \cite{calico}, Istio \cite{istio}, Bastion \cite{bastion}, and eZTrust \cite{eztrust} regulate communication at individual requests or endpoints. These approaches scale well but cannot capture multi-hop dependencies.

    \item \textit{Information-flow defenses.} Trapeze \cite{trapeze} and Fabric \cite{fabric} extend visibility across hops but ignore hop ordering, a distinction essential for separating benign workflows from malicious ones.

    \item \textit{Multi-hop enforcement.} CloudCover~\cite{cloudcover}, Grewal et al.~\cite{expressive}, Meadows et al.~\cite{sidecar}, and WILL.IAM~\cite{william} enforce request ordering but
    route traffic through a centralized controller for decision making,
    which scales poorly, adds high latency, and creates a single point of failure.
\end{itemize}

In contrast, \tool is the first system to simultaneously provide multi-hop enforcement, hop-ordering preservation, and decentralized scalability.
It advances the semantics of microservice security policies beyond prior single-hop access control mechanisms~\cite{cilium, calico, istio, bastion, eztrust} and information-flow defenses that omit request ordering~\cite{trapeze, fabric}. 
Compared to early multi-hop access control systems~\cite{cloudcover, expressive, sidecar, william, kalium}, \toolnospace's primary contribution is not new policy expressiveness, but scalable, transparent, decentralized enforcement.

\myparagraph{History-based access control extensions}
Access control models that incorporate historical context have been studied. The Chinese Wall security policy~\cite{36295} is an early example, restricting a subject’s future accesses based on prior actions to avoid conflicts of interest. Subsequent research extended history-based models to domains like software security, where the execution stack serves as history~\cite{fournet2003access}, and mobile application security, where system events provide context~\cite{10.1145/288090.288102}. Fong~\cite{1301314} later formalized a general history-based framework.

In this work, we extend history-based access control to microservices, with three components:

\textbf{(i) Requests.} 
In microservices, the fundamental events to monitor and enforce are requests. Requests typically represent service-to-service communication but the concept extends to other granularities such as pods, IP ranges, or nodes. Similarly, requests may be issued over different protocols (e.g., HTTP, RPC). Thus, a request can be broadly defined as an interaction between two entities in the system, independent of the specific abstraction or protocol.

\textbf{(ii) Causal dependencies.}
Requests are often causally related, as one request may trigger others to fulfill users' intent, a common pattern that have been widely studied \cite{meta,uber_ds,ali_ds,bd_ds}. Dependencies can be synchronous or asynchronous.
In synchronous dependencies, request $r_1$ triggers request $r_2$ and cannot complete until $r_2$ finishes, a pattern common in aggregator services.
In asynchronous dependencies, $r_1$ triggers $r_2$ but may complete independently, as in queuing systems such as RabbitMQ~\cite{rabbitmq} and Kafka~\cite{kafka}.

\textbf{(iii) Request workflows as history.}
A request workflow is defined as a sequence of requests in which each request is causally dependent on its predecessor. In practice, the set of valid workflows within a microservice application is finite, since it is constrained by the architecture and service interactions. This set is also substantially smaller than the space of all possible service permutations, making it a practical basis for access control. In our history-based access control model, workflows form the historical context: \emph{a request is permitted only if both the request itself and the causal sequence leading to it are valid}. This ensures that even individually valid requests are denied if they deviate from legitimate workflows.

\begin{figure*}[h]
    \centering
    \includegraphics[width=\textwidth]{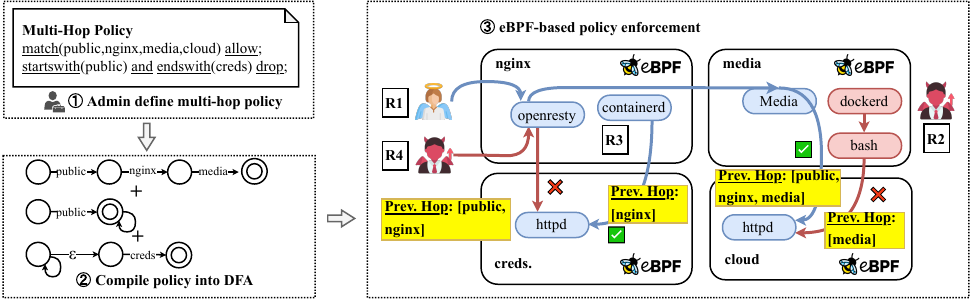}
    \caption{Overview of \tool. Administrators define multi-hop policies (\textcircled{1}, \cref{sec:policy}), which are compiled into a single DFA for efficient label propagation and real-time enforcement (\textcircled{2}, \cref{sec:encoding}) and enforced in a distributed manner using eBPF (\textcircled{3}, \cref{sec:tracing}). \tool blocks multi-hop attacks that compromise services (\fbox{R2}) or leverage benign services (\fbox{R4}) without disrupting legitimate workflows (\fbox{R1}, \fbox{R3}). For example, both \fbox{R1} and \fbox{R2} send requests from \texttt{media} to \texttt{cloud}, but only the former follows the intended workflow; similarly, both \fbox{R3} and \fbox{R4} send requests from \texttt{nginx} to \texttt{creds}, but the latter is not triggered by a valid request path.}
    \label{fig:overview}
\end{figure*}

\subsection{Threat Model}
\label{subsec:threat_model}
Our work focuses on multi-hop attacks.
\tool is designed to be deployable in both self-managed microservice clusters to enforce application-specific request workflows, and multi-tenant microservice clusters to enforce per-tenant workflows and prevent cross-tenant attacks.
We consider attackers who leverage legitimate inter-service requests as stepping stones to access protected services.
Such attacks may originate externally through public-facing endpoints or internally through malicious container images.
Attackers may perform multi-hop attacks by abusing benign microservice functionality (e.g., server-side request forgery, open-redirection vulnerabilities) or by spawning malicious processes after compromising a service.
We assume the underlying platform, namely the host kernel, the eBPF subsystem, and the container runtime, remains trustworthy, so attackers cannot subvert the kernel or the execution environment of eMicro's eBPF programs.
Adversaries with kernel-level access could disable the microservice platform entirely, which is beyond the scope of this work and falls under kernel security rather than microservice security.
This threat model is aligned with prior work on cloud and network security~\cite{cloudcover,kalium,trapeze,expressive,william}.


\section{System Overview}

\tool protects microservices from multi-hop attacks. 
Instead of relying on centralized enforcement that cannot scale, \tool provides scalable, distributed enforcement with history-based access control.
It realizes this through \textcircled{1} an expressive multi-hop policy language, \textcircled{2} compact deterministic finite automaton (DFA) encoding of request sequences, and \textcircled{3} transparent, application-agnostic, in-kernel enforcement using eBPF.

\cref{fig:overview} presents a system overview.
In step \circled{1}, administrators define multi-hop security policies using {\toolnospace's} policy language (\cref{sec:policy}), specifying valid request workflows over complete request paths.
Typical use cases include preventing unauthorized access, blocking credential leakage, and enforcing control-flow integrity.

Once policies are defined, \tool compiles them into a DFA (step \circled{2}, \cref{sec:encoding}). Each atomic predicate is translated into a DFA, while composite predicates are constructed using union and intersection operations. DFAs from multiple policies are then merged under deny-by-default semantics using a data-parallel algorithm~\cite{data_parallel} that combines large policy sets efficiently. The resulting DFA is minimized using Hopcroft’s algorithm~\cite{hopcroft}. \emph{Each DFA state corresponds to a valid request sequence} and is assigned a unique label. Requests carry only this label.

In step \circled{3}, \tool enforces multi-hop policies in a fully distributed and application-agnostic manner (\cref{sec:tracing}).
Each request workflow is assigned an initial DFA label based on its entry point.
As requests propagate across services, \tool observes network packets and execution contexts at the kernel level and transparently propagates and updates labels across packet headers, processes, threads, and coroutines. 
At each hop, label is validated against the DFA transition function and updated to reflect the extended request history. Requests whose labels reach an accepting state continue execution, while those that violate the DFA are blocked. 
Propagation and enforcement are implemented entirely in the kernel using eBPF-based tracing and execute independently on each service node, eliminating central bottlenecks and enabling low-latency, constant-time checks. 
To support fast distributed enforcement, the DFA and state mappings are stored in BPF maps and replicated across nodes using a consensus protocol Raft~\cite{raft}, allowing local enforcement without additional network communication.


\section{System Design}

\subsection{Multi-Hop Policy Language}\label{sec:policy}

To prevent multi-hop attacks, \toolnospace's policy language provides predicates for pattern matching over complete request paths. As shown in \cref{fig:policylang}, atomic predicates (\texttt{match}, \texttt{startswith}, \texttt{endswith}, and \texttt{contains}) define regular languages over request paths \texttt{PH}, while predicate composition using \texttt{and}, \texttt{or}, and \texttt{not} corresponds to language intersection, union, and complement. A policy associates a predicate with an action. Enforcement actions (\texttt{allow}, \texttt{drop}) determine whether a request path is accepted or rejected, whereas auxiliary actions (\texttt{inspect}, \texttt{alert}, \texttt{reset\_label}, and \texttt{set\_label(x)}) introduce side effects without changing path acceptance. 

We illustrate the \toolnospace's expressiveness through three multi-hop policies that capture common security requirements:

\textbf{(i) Unauthorized access prevention.}
The administrator requires that every public request to the \texttt{cloud} database be properly authorized through the public gateway \texttt{nginx}; any request that bypasses \texttt{nginx} (for example, via an accidentally exposed service port) must be dropped.
This can be encoded as: \texttt{\underline{startswith}(public) \underline{and} \underline{not} \underline{contains}(nginx) and \underline{endswith}(cloud) \underline{drop}}.
Because the policy reasons over the full request workflow, it catches violations even when \texttt{nginx} is not the immediate predecessor of \texttt{cloud}.

\textbf{(ii) Credentials leakage prevention.}
Administrators require the \texttt{creds} service to act as a terminal service that never initiates outbound requests, especially those reaching \texttt{public}.
This can be encoded as
\texttt{\underline{startswith}(creds) \underline{and} \underline{endswith}(public) \underline{drop}}.

\textbf{(iii) Control flow integrity.}
Administrators want to enforce exact request flows by enumerating all valid paths (including control-flow variations) and disallowing the rest. The example in \cref{fig:overview} can be implemented as
\texttt{\underline{match}(public, nginx, media, cloud) \underline{allow}; \underline{match}(nginx, creds) \underline{allow}}.

\begin{figure}[t]
\centering    
\small
\[
\begin{array}{rcl}
\langle Policy \rangle   &::=&  \langle Predicate \rangle\ \langle Action \rangle \\[2pt]
\langle Action \rangle   &::=& \texttt{drop}\ \mid\ \texttt{allow} \mid\ \texttt{alert} \mid\ \texttt{inspect} \mid\ \\ &&  \texttt{reset\_label} \mid\ \texttt{set\_label(x)} \\[2pt]
\langle Predicate \rangle&::=& \texttt{match(}\langle PH\rangle\texttt{)} \\
                         &\mid& \texttt{startswith(}\langle PH\rangle\texttt{)} \\
                         &\mid& \texttt{endswith(}\langle PH\rangle\texttt{)} \\
                         &\mid& \texttt{contains(}\langle PH\rangle\texttt{)} \\
                         &\mid& \langle Predicate \rangle\ \langle Op \rangle\ \langle Predicate \rangle \\[2pt]
                         &\mid& \texttt{not} \langle Predicate \rangle\ \\[2pt]
\langle Op \rangle       &::=& \texttt{and}\ \mid\ \texttt{or} \\[2pt]
\langle PH \rangle       &::=& \texttt{service*}
\end{array}
\]
\caption{Multi-hop microservices policy language}
\label{fig:policylang}
\end{figure}

\subsection{Compiling Policies into DFA} \label{sec:encoding}

Multi-hop access control requires propagating request-history context, but naively carrying full request sequences and sequentially validating them against policies does not scale.

\myparagraph{Naive encoding and its inefficiency}
The simplest encoding is to represent the hop sequence as a list of integers.
For instance, the workflow \verb|A,B| can be represented as the list \verb|[1,2]|, and the workflow \verb|A,B,C| as \verb|[1,2,3]|. 
The list must accommodate up to $m$ elements (the maximum depth of any request workflow) and each element must uniquely identify $n$ services.
Under this design, the required bits can be derived as follows:
\begin{equation}
    \label{eq:naive}
    b_{naive} = \log_{2}(n) \times m
\end{equation}
where $b_{naive}$ denotes the label size in bits, $n$ denotes the number of unique services, and $m$ denotes the maximum request workflow's depth. 

Label size is critical as labels are carried with the requests across thousands of services and multiple hops.
Inefficient designs can cause unnecessary space overhead. 
Note that the encoding only needs to represent valid request workflows, since invalid ones are discarded upon sending a malicious request. 
In contrast, a naive scheme encodes \emph{all} permutations of $n$ services of length $m$ and most of them are invalid request workflows.

\myparagraph{DFA encoding}
To propagate request sequences efficiently, \tool assigns each valid request sequence a unique label. 
For example, \verb|[A,B]| is labeled \verb|1|, while \verb|[A,B,C]| is labeled \verb|2|, ensuring that no labels are wasted on invalid workflows. 
We formalize this approach with a DFA, where each valid sequence corresponds to a state. \emph{This compact representation minimizes the number of bits required for label storage}, as follows:
%
\begin{equation}
    \label{eq:dfa}
    b_{dfa} = \log_{2}(s)
\end{equation}
where $b_{dfa}=$ tag size in bits and $s=$ number of states.

$b_{dfa}$ is substantially smaller than $b_{naive}$ as $s << n^m$ in most cases.
In the worst case scenario, where all sequences from length of $1$ until $m$ are valid, the value of $s$ is as follow:
\begin{equation}
    \label{eq:dfa_upper_bound}
    s_{max} =\sum_{k=1}^{m}n^{k}=n\frac{n^m-1}{n-1}
\end{equation}
Here, $s$ exceeds $n^m$, but the bit difference ($\Delta$) is negligible.
\begin{equation}
    \label{eq:bit_difference}
    \begin{aligned}
    \Delta
    &= \log_2(s_{max}) - m \log_2(n) \\
    &= [\log_2(n) - \log_2(n-1)]
     - [\log_2(n^m) - \log_2(n^m - 1)]
    \end{aligned}
\end{equation}
Even under extreme conditions (e.g., $n=2$, $m=2$), the difference is only 1 bit and shrinks as $n$ and $m$ grow. Therefore, DFA encoding is at worst negligibly less efficient by one bit and more compact than naive encoding most of the time.

\myparagraph{Compilation process}
The DFA describes the valid request workflows, which is derived from the multi-hop policies.
\tool compiles policies into a single DFA through two components.
First, a left-to-right parser converts policies into an abstract syntax tree (AST).
Second, the AST passes collect all participating services, convert each atomic predicate into a DFA, and merge them into a single DFA.

Policies are transformed into a DFA as follows.
Each predicate \texttt{p} denotes a regular language and is therefore expressible as a DFA that accepts a language $\mathcal{L}$ according to Kleene’s theorem~\cite{kleene}.
The composition \verb|p1 or p2| corresponds to the union $\mathcal{L}' = \mathcal{L}_1 \cup \mathcal{L}_2$, while \verb|p1 and p2| corresponds to the intersection $\mathcal{L}' = \mathcal{L}_1 \cap \mathcal{L}_2$.
Actions determine the final DFA semantics: the policy \verb|p allow| corresponds to $\mathcal{L}$ itself, whereas \verb|p deny| corresponds to its complement $\mathcal{L}' = \overline{\mathcal{L}}$.

\begin{figure}[t]
    \centering
    \begin{subfigure}[t]{\linewidth}
        \centering
        \includegraphics[width=.95\linewidth]{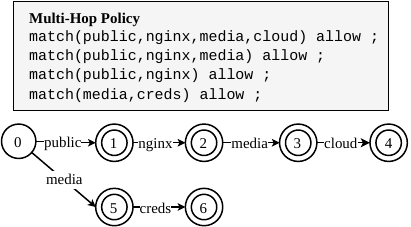}
        \caption{DFA representation for a multi-hop policy}\label{fig:dfa_illustration}
    \end{subfigure}

    \vspace{2.5em}  
    
    \begin{subfigure}[t]{\linewidth}
        \centering
        \includegraphics[width=.95\linewidth]{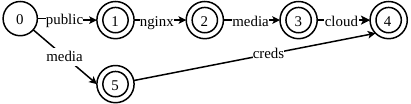}
        \caption{Minimized DFA}\label{fig:minimized_dfa}
    \end{subfigure}

    \caption{DFA encoding for multi-hop policies. A six-service deployment with the given set of valid request workflows requires only eight labels to uniquely encode each valid sequence, reduced to five after minimization.}
\end{figure}

Multiple policies are \emph{merged} into a single DFA as follows.

\begin{equation} 
    \Biggl(\;\bigcup_{i=1}^{n} \mathcal{L}_{a,i}\;\Biggr) \;\cap\; \Biggl(\;\bigcap_{j=1}^{n} \mathcal{L}_{d,j}\;\Biggr) 
\end{equation}

\noindent where $\mathcal{L}_{a,i}$ denotes the $i$-th allow policy and $\mathcal{L}_{d,j}$ denotes the $j$-th deny policy.
This construction requires a request to satisfy at least one allow policy while violating none of the deny policies, thereby enforcing deny-by-default semantics.
The multi-hop policies and compiled DFA are illustrated in \cref{fig:dfa_illustration}.

The union and intersection operations on two DFAs have a time complexity of $\mathcal{O}(m \cdot n \cdot k)$, where $m$ and $n$ are the numbers of states in the two DFAs and $k$ is the alphabet size.
Our initial experiments show that compiling 1,000 policies into a single DFA takes more than 20 minutes.
Hence, \tool applies optimization.
In general, union and intersection operations on one pair of DFAs are independent of those on other pairs.
These operations are both commutative and associative, making them parallelizable.
\tool leverages a data-parallel algorithm~\cite{data_parallel} such that each process produces a partial result, reducing the number of DFAs by half at each iteration, until a single DFA remains.

To keep the DFA size minimal in the presence of thousands of services and request workflows, we apply \emph{DFA minimization}, which is a classical problem in automata theory, typically solved using Hopcroft's algorithm~\cite{hopcroft}, shown in \cref{alg:dfa-min}.
The algorithm groups indistinguishable states that have identical transition behavior by partitioning states such that those in the same partition are equivalent.  
It begins with two coarse partitions, accepting and non-accepting states, and iteratively refines them until no further refinement is possible. 
\tool leverages this algorithm by adapting the definitions of states and the transition function to its own model.
The minimized DFA, after merging non-distinguishable states, is illustrated in \cref{fig:minimized_dfa}.

\myparagraph{Extending DFA for advanced use case}
Although inter-service call cycles are uncommon in conventional microservice architectures, they can arise in practice. Such behaviors do not violate the DFA-based model: bounded service-call loops unroll into finite sequences, while unbounded iterative loops correspond to cycles in the state graph and remain regular by closure of regular languages under Kleene star. 
If future microservice workflows require unbounded recursion, a more expressive formalism, such as a pushdown automaton, could be used.

\subsection{Label Propagation and Policy Enforcement} \label{sec:tracing}

\tool propagates request labels encoding requests' previous hops as DFA states within and across services to enforce multi-hop policies. 
A \emph{naive} approach is to attach them to application-level requests (e.g., HTTP headers in HTTP requests, gRPC metadata in RPC), as seen in previous studies and tools (CloudCover~\cite{cloudcover}, Grewal et al.~\cite{expressive}, Meadows et al.~\cite{sidecar}, Jaeger~\cite{jaeger}).
Handlers can then parse and propagate labels through dedicated or instrumented functions. 
However, microservices are implemented in diverse environments.
For example, Meta reports 16 languages in production \cite{meta}, while Jaeger and OpenTelemetry \cite{jaeger} maintain instrumentation across 11 languages.
Beyond languages, organizations adopt custom dialects (e.g., Meta’s Hack, a PHP variant \cite{meta}) and a variety of libraries and frameworks. 
Manually instrumenting each of them is difficult and error-prone, especially under frequent updates, motivating a general solution.

\begin{algorithm}[t]
    \caption{Hopcroft Algorithm \cite{hopcroft}}
    \begin{algorithmic}[1]
        \State $P \gets \{F, Q \backslash F\}$
        \Comment{Current partition}
        \State $W \gets \{F, Q \backslash F\}$
        \Comment{Distinguisher candidate}
        \While {$W$ is not empty}
            \State Choose $A \in W$
            \State $W \gets W \backslash \{A\}$
            \ForAll {$c \in \Sigma$}
                \State $X \gets \{x | x \in Q , \delta[x][c] \in A\}$
                \ForAll {$Y \in P$}
                    \If {$X \cap Y = \emptyset | Y \backslash X = \emptyset$}
                        continue
                    \EndIf
                    \If{$Y \in W$}
                    \Comment {$Y$ has not been used}
                        \State $W \gets W \backslash {Y}$
                        \State $W \gets W \cup \{X \cap Y, Y \backslash X\}$
                    \Else
                        \If {$|X \cap Y| \leq |Y \backslash X|$}
                            \State $W \gets W \cup \{X \cup Y\}$
                        \Else 
                            \State $W \gets W \cup \{Y \backslash X\}$
                        \EndIf
                    \EndIf
                \EndFor
            \EndFor
        \EndWhile
    \end{algorithmic}
    \label{alg:dfa-min}
\end{algorithm}

To address this challenge, \tool adopts an in-kernel approach that remains \emph{agnostic} to specific languages, frameworks, libraries, and application-level implementations.
The kernel does not directly observe application-level abstractions; instead, it views executing programs as processes and threads, and application-level requests (e.g., HTTP or RPC) as network packets.
\tool propagates labels along these kernel-visible entities.
For network packets, \tool \emph{attaches a custom header after the L4 header}, preserving compatibility with existing L1--L4 functionality such as routing and firewalls.
For processes, threads, and coroutines, \tool taints them with labels stored in BPF maps, keyed by PID, TID, or coroutine context.
Beyond agnosticism, this kernel-level design also improves performance by avoiding costly user/kernel context switches.

\myparagraph{eBPF-based tracing}
Directly modifying the kernel source code is cumbersome and error-prone.
\tool instead leverages eBPF~\cite{ebpf}, a kernel technology that allows custom programs to run efficiently and safely inside the kernel.
Originally designed for packet filtering, eBPF has been extended to support observability, security monitoring and defenses~\cite{upolicycraft,tetragon,bajaber2026netcap},and network performance optimization~\cite{katran}. 
eBPF programs are attached to various hook points in the kernel, such as traffic control (TC), express data path (XDP), and kprobes, allowing users to intercept and run custom logic alongside kernel code.

To propagate labels across kernel-visible entities, \tool leverages three abstractions: network packets, sockets, and execution units. 
Sockets send and receive packets. Execution units create sockets, consume messages from them, and transmit messages through them. 
An execution unit is the smallest unit of a program in execution that processes a single request at a time (e.g., a process, thread, or coroutine).

To capture interactions among these entities, \tool attaches five eBPF programs to various kernel hooks as shown in \cref{fig:tracing}. 
Programs \circled{a} and \circled{e} are attached to each service’s virtual network interface to propagate labels across services. 
Programs \circled{b}, \circled{c}, and \circled{d} are attached to the host kernel, which is shared by all services on that node, to propagate labels within a service. 
The host kernel is shared between services and it retains visibility into execution units and sockets across all services despite namespace isolation. 
We describe each eBPF program below:

\begin{figure}
    \centering
    \includegraphics[width=\linewidth]{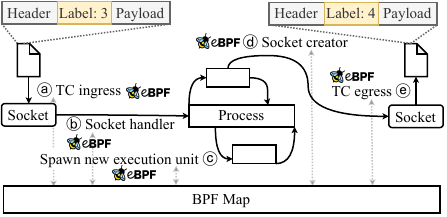}
    \caption{In-kernel tracing with eBPF. Five hooks are used to propagate labels throughout the request workflows.}
    \label{fig:tracing}
    \vspace{-1ex}
\end{figure}

\textbf{\circled{a} TC ingress hook.}
This program runs on each packet arriving at the virtual network interface, parsing the packet header to extract the label, which is present only in the first packet of a flow. 
Subsequent packets omit the label because they belong to the same request. 
The program then advances the label to the next state based on the DFA transition function.
If the transition is valid, the packet is accepted; otherwise, it is dropped.
The receiving socket, identified by its five-tuple, is then tagged with the updated label.
Such labels are stored in a BPF map that maps each kernel-visible entity to its current label.

\textbf{\circled{b} Socket handler.} 
This eBPF program captures the event when an execution unit consumes messages from a socket. 
On Linux kernel 5.15, it is attached as a kernel return probe (\verb|kretprobe|) on the \verb|inet_csk_accept| function. 
The program inspects the internal kernel socket structure to check whether a socket carries a label, and if so, propagates that label to the consuming execution unit.

\textbf{\circled{c} Spawn new execution unit.} This eBPF program captures the event when an execution unit spawns a new one. For multi-process or multi-thread systems, this program is attached as a kernel return probe (\verb|kretprobe|) on the \verb|clone| syscall. 
For concurrent systems where the execution unit is a coroutine, this program is attached as a user-space probe (\verb|uprobe|) on the runtime's coroutine-scheduling function (e.g., \verb|EventLoop| class in Python's asyncio, or \verb|runtime.casgstatus|
  in Go's goroutine runtime)

\textbf{\circled{d} Socket creator.} 
This eBPF program captures the event when an execution unit initiates a new request to another service. 
On Linux kernel 5.15, it is attached as a kernel probe (\verb|probe|) on the \verb|tcp_connect| function. 
This program checks whether the execution unit is already associated with a label. 
If yes, it propagates the label to the related socket that will send the request. 

\textbf{\circled{e} TC egress hook.}
This program is triggered whenever a packet leaves a virtual network interface. 
It checks whether the socket associated with this request carries a label; if so, it attaches the label as a custom header in the packet. 
Otherwise, it creates a new label representing a single-hop request workflow consisting of the current service.

In deployments with multiple containers on the same physical node, programs \circled{a} and \circled{e} run as one instance per service, each attached to that service's virtual network interface.
\tool attaches these programs at service startup and maintains a mapping between network interface identifiers and labels. 
The host-kernel programs (\circled{b}, \circled{c}, \circled{d}) run as a single shared instance per node and observe all containers and processes on that host. Although each container has its own PID namespace, the host kernel sees all processes under globally unique PIDs. 
\tool combines container (cgroup) IDs with PIDs to distinguish processes within each microservice.

\myparagraph{Policy enforcement and distribution}
\tool stores the multi-hop policy, encoded as a DFA, in a BPF map.
BPF map is an in-kernel key-value store that naturally represents the DFA transition function: \verb|(previous_state, next_hop)| serves as the key and \verb|(next_state)| as the value.
For example, the BPF map for \cref{fig:minimized_dfa} maps \texttt{(0, public)} to value \texttt{1} and \texttt{(0, media)} to value \texttt{5}.
This design provides high-performance persistent states shared across eBPF programs and lets each service node consult its local map directly, without extra network requests, enabling distributed and decentralized enforcement.
The map can also be updated by user-space programs without recompiling or reloading eBPF code, simplifying policy updates.
As a result, \emph{each access control decision becomes a constant-time BPF map lookup}: a request to \texttt{svc\_a} carrying label \texttt{l} is allowed if and only if the BPF map contains a valid transition from \texttt{(l, svc\_a)} to some next state \texttt{l'}.

\tool distributes policies to participating nodes and maintains consistency using the Raft consensus algorithm~\cite{raft}.
Policy updates are propagated across nodes and committed once a majority of nodes accept them.
Failed nodes restart and recover state up to the latest committed update.
Nodes exchange periodic heartbeats to ensure leader liveness and to detect and recover missing updates.
While policies are replicated across physical nodes, enforcement remains local without additional network overhead.


\section{Security Analysis}\label{sec:security_analysis}

We analyze the security of \tool under the threat model in \cref{subsec:threat_model}.
\tool remains resilient against adaptive attackers who know its design and actively attempt to evade, disable, or overload policy enforcement.
By enforcing multi-hop policies entirely in kernel space with bounded overhead, \tool prevents tracing evasion, detects replayed request histories, and withstands denial-of-service attacks.

\myparagraph{Evading \tool tracing}
An attacker who compromises a service may attempt to evade \tool by spawning new processes or threads (e.g., launching a bash shell, escalating privileges, or executing malicious scripts).
These attempts fail because \tool propagates labels across execution units: child execution units inherit the parent's label.
An attacker may further try to escape enforcement by creating a seemingly untainted execution unit with no label.
This is not possible: every execution unit has a parent, and if the parent carries no prior history, the child is assigned the initial label \texttt{0}, representing a locally initiated request.
This does not grant unrestricted privileges: whether requests carrying label \texttt{0} are permitted is determined by the administrator's policy, and access is granted only if the resulting request path matches an allowed workflow. While an attacker may still invoke workflows explicitly permitted to originate locally (e.g., routine tasks), they remain confined to that predefined scope and cannot exploit multi-hop interactions to reach unauthorized paths.

\myparagraph{Store-and-forward}
An attacker may attempt to chain multiple individually valid workflows to bypass the policy.
%
For example, an attacker may bend legitimate workflows through vulnerabilities such as remote code execution or open redirection, or by exploiting permissive suffixes in allowed paths.
\tool defeats these because each request carries its full causal lineage: even if the bent service is itself allowed to perform certain calls, the resulting end-to-end path (e.g., from an internal sensitive service to an untrusted destination) does not match any administrator-defined workflow and is rejected.
More advanced attackers may use indirect store-and-forward: data intended for exfiltration is first staged in an intermediate service and later retrieved through a separate workflow.
\tool can be extended with additional in-kernel eBPF probes to trace such data-mediated dependencies.
For example, filesystem-mediated chains can be reconnected by tracing \texttt{inode} operations: by observing which execution units create, write, and consume each \texttt{inode}, \tool propagates labels through the file dependencies and reconstructs the causal chain.
We implement this by extending the socket tracing programs (\circled{b} and \circled{e} in \cref{fig:tracing}) from socket file descriptors to user-owned files, attaching probes to the kernel functions \texttt{vfs\_create}, \texttt{vfs\_write}, and \texttt{vfs\_read} to tag the corresponding \texttt{inode}.

\myparagraph{Replay attacks}
An attacker may attempt to reuse labels from legitimate request workflows, for example, by eavesdropping on traffic or performing man-in-the-middle attacks.
\tool prevents such replays by binding labels to the execution and communication context in which they were generated: it traces parent-child relationships among execution units and data flows between execution units and network packets, capturing who spawns whom, who sends data, and who receives it.
If an attacker injects a valid label into a different execution context, the label becomes inconsistent with the observed execution and communication history, and the request is rejected during policy evaluation.

\myparagraph{Denial-of-service attacks}
An attacker may attempt to disable \tool by flooding it with requests.
\tool is resilient to such attacks because policy enforcement runs entirely in kernel space and requires no extra network communication. Each policy check executes in constant time, independent of request history length.
Thus, \tool does not introduce a new bottleneck in the networking stack or virtual interfaces.
In practice, any denial-of-service attack sufficient to overwhelm \tool would also overwhelm co-located microservices, rather than selectively disabling enforcement while leaving services unprotected.


\section{Evaluation} \label{sec:eval}

We evaluate \tool along three dimensions: (i) effectiveness in preventing multi-hop attacks, (ii) runtime efficiency under realistic microservice workloads, and (iii) scalability of its policy encoding and in-kernel enforcement mechanisms.
\tool is implemented in approximately 700 lines of C and Python.
We conduct end-to-end experiments on a controlled microservice cluster consisting of two physical machines hosting nine virtual machines.
Each physical machine is equipped with an Intel Core i7-13700 CPU (2.1\,GHz), 32\,GB RAM, and runs Ubuntu~20.04.
This setup allows us to directly measure enforcement overhead and the impact of in-kernel tracing under concurrent request execution.

Each VM runs the social network application from DeathStarBench~\cite{deathstarbench}, comprising 27 microservices with 10 replicas, 64 inter-service communication edges, and 32 distinct request workflows.
We deploy the application using the benchmark's provided Docker images, Docker Compose configuration, and Kubernetes deployment scripts, ensuring fidelity to widely used microservice development and deployment practices.
We scale the workload generator to issue a total of 12 million requests under sustained concurrency, exercising overlapping request workflows and stressing policy enforcement under load.
In addition to benign workloads, we explicitly construct and inject representative multi-hop attacks drawn from real-world attack patterns into the running application to evaluate enforcement correctness under adversarial conditions.

To evaluate \tool under diverse networking conditions encountered in modern cloud environments, we repeat experiments across multiple container networking configurations, including Linux bridge networking, Cilium~\cite{cilium}, Flannel, and Calico~\cite{calico}.
Beyond online experiments, we evaluate the scalability of \toolnospace's DFA-based policy encoding using production-grade distributed tracing datasets from Uber~\cite{uber_ds}, Alibaba~\cite{ali_ds,ali_ds2}, and ByteDance~\cite{bd_ds}.
These datasets contain up to 1.4 million traces collected over seven days, totaling 570\,GB of data (shown in \cref{tab:dataset}).
Each trace is recorded in Jaeger format~\cite{jaeger} and captures service-to-service requests linked via explicit parent-child relationships.
We reconstruct request workflows from these traces to quantify label space requirements and assess the scalability of \toolnospace's DFA encoding under realistic cloud-scale service topologies.

Our evaluation answers five key research questions.

\begin{enumerate}[label={\textbf{(RQ\arabic*)}}, leftmargin=*,itemsep=6pt, topsep=6pt]
    \item \textit{How effective is \tool at protecting microservices from multi-hop attacks?}
    We evaluate \toolnospace's effectiveness in blocking multi-hop attacks, including path-suffix and workflow-bending attacks, and compare its policy coverage over potential attack paths to that of adjacent-hop access control.

    \item \textit{How efficient is \tool at enforcing multi-hop access control in microservices?}
    We measure end-to-end throughput and request workflow latency, along with the cost of policy compilation and deployment, comparing against prior centralized enforcement approaches.
    
    \item \textit{How much space does \toolnospace’s DFA-based encoding save compared to naive encoding?}
    We quantify label-space overhead by comparing naive request-history encoding with \toolnospace's DFA-based encoding on production-scale cloud traces.

    \item \textit{How scalable are \toolnospace's in-kernel enforcement and label propagation?}
    We measure the overhead of eBPF-based request tracing, along with policy lookup latency and throughput across hundreds to millions of policies.

    \item \textit{How much overhead does \tool incur to maintain consistency and availability across nodes?}
    We evaluate \toolnospace's ability to support consistent distributed policy updates and to maintain correct enforcement under node and leader failures.
    
\end{enumerate}

\subsection{RQ1. Defense Effectiveness}
\label{subsec:rq1}

\begin{figure*}[]
    \centering
    \includegraphics[width=.9\textwidth]{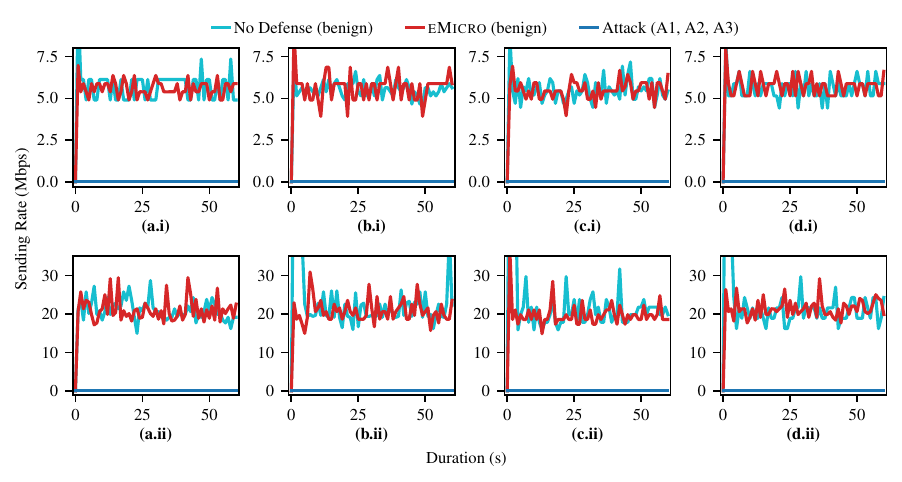}
    \vspace{-1ex}
    \caption{Service-to-service TCP sending rate over time under benign workloads and representative multi-hop attacks (A1: workflow bending via redirection, A2: RCE-driven lateral movement, A3: malicious container image enumeration). Results are shown across four container networking configurations: (a) Linux bridge, (b) Cilium, (c) Flannel, and (d) Calico. For each configuration, we evaluate two WAN settings: (i) 10 ms latency, 1 ms jitter, 0.1\% packet loss, and (ii) 20 ms latency, 5 ms jitter, 0\% packet loss.}
    \label{fig:sending_rate}
\end{figure*}

We evaluate \toolnospace's defense effectiveness against three representative classes of multi-hop attacks.
\emph{First (A1)}, we exploit an intentionally introduced open redirection vulnerability in a service without fully compromising it. This attack coerces the vulnerable service into issuing requests to a downstream service that is valid for an adjacent hop but invalid over the complete request path, bending the intended request workflow.
\emph{Second (A2)}, we exploit an injected remote code execution (RCE) vulnerability in a service, escalate privileges, and initiate connections to other services via SSH.
Each individual hop is permitted under adjacent-hop access control, but the resulting end-to-end path violates the intended request workflow.
\emph{Third (A3)}, we exploit a misconfigured Docker daemon that grants non-privileged users access to the Docker API, allowing them to launch malicious services via \texttt{docker run} or \texttt{docker exec}.
The malicious service performs network enumeration and issues requests that follow valid request-workflow suffixes.
However, because they originate locally from the malicious service rather than from the correct preceding hops, they carry an incorrect context (initial label \texttt{0}) instead of a valid state propagated from prior hops, violating the intended request workflow.

We evaluate across four container networking configurations (Linux bridge, Cilium, Flannel, and Calico) and two WAN conditions with latency, jitter, and packet loss of (10 ms, 1 ms, 0.1\%) and (20 ms, 5 ms, 0\%), yielding eight experimental scenarios.
For each attack scenario, we measure under three configurations: (i) a no-defense baseline, (ii) \tool enabled under benign workloads, and (iii) \tool enabled with the attack active.
Defense effectiveness is measured by the number of malicious requests dropped.
Performance is measured by the service-to-service TCP sending rate (using \texttt{iperf}) since enforcement runs on the network packet path. A sending rate of zero means requests are being dropped.

The results consistently show that \tool drops all multi-hop attacks (A1, A2, A3) while preserving all benign requests.
It maintains traffic performance, introducing less than 1\% reduction in service-to-service TCP sending rate relative to the no-defense baseline (\cref{fig:sending_rate}); the impact is not statistically significant.

\myparagraph{Multi-hop policy coverage}
Beyond blocking concrete attacks, we measure how effectively multi-hop policies reduce the attack surface compared to adjacent-hop access control.

We use the first two days of the Alibaba~\cite{ali_ds2,ali_ds} and Uber~\cite{uber_ds} traces and reconstruct two types of workflows of lengths 2, 4, and 6.
For valid workflows, we recover the causal request chains actually recorded in the dataset by chain requests via their unique identifiers.
For invalid workflows, we chain request pairs by matching the destination of one request to the source of the next: each individual hop is valid, yet the chain itself is invalid, indicating the potential attack paths an adversary can exploit.
We then measure \emph{policy coverage}, defined as the fraction of all workflows each access control mechanism classifies correctly, i.e., valid workflows permitted and invalid workflows denied.

\cref{tab:attack_surface} shows that \tool maintains 100\% coverage at every workflow length, permitting valid workflows and denying invalid ones, because enforcement is bound to the full request history.
Adjacent-hop access control, in contrast, inspects only immediate hops and retains no history, so it admits any chained workflow whose individual hops are independently allowed, but the entire path can be malicious.
Its coverage therefore degrades as workflows grow longer and the space of attacker-reachable chains expands: at length 6, adjacent-hop correctly classifies only 0.3\% of feasible workflows.
The gap in attack-surface coverage between the two mechanisms grows from $1.01\times$ (2-hop) to $1000\times$ (6-hop), with a geometric mean of $40\times$.

\myparagraph{Resilience against store-and-forward evasion}
We evaluate a store-and-forward attack in which two individually valid workflows are chained into an invalid end-to-end workflow, by transferring a secret token file to an intermediate service, storing it, and later consuming it through another valid workflow. 
With data tainting enabled (\cref{sec:security_analysis}), \tool blocks all such attempts by propagating labels through inode creation, writes, and reads, so the chain remains visible even when data is persisted to a file.
Overhead is minimal: each tracing entry occupies an 8-byte key plus a 2-byte (65K states) or 3-byte (16M states) DFA label, and each tracing operation, including inode reads and writes, incurs less than 2 $\mu$s.

\begin{tcolorbox}[colback=magenta!5, colframe=magenta!80!black,left=3pt,right=3pt,top=3pt,bottom=3pt]
\textbf{Takeaway:} \tool blocks all multi-hop attacks, covering 40$\times$ more attack paths on average than adjacent-hop access control, with minimal performance impact on traffic.
\end{tcolorbox}

\begin{table}[t]
    \centering
    \small
    \caption{\toolnospace's multi-hop policies coverage}
    \begin{tabular}{lccc}
        \toprule
        \textbf{Dataset} & \textbf{\# Hops} &  \textbf{Adj. Hop (\%)} & \textbf{\tool (\%)} \\ 
        \midrule
        \multirow{3}{*}{Alibaba \cite{ali_ds,ali_ds2}} & 2 & 99.9 & 100.0 \\
                                   & 4 & 15.6 & 100.0 \\
                                   & 6 & 0.3 & 100.0 \\
        \midrule
         \multirow{3}{*}{Uber \cite{uber_ds}} & 2 & 2.7 & 100.0 \\
                                   & 4 & 1.2 & 100.0 \\
                                   & 6 & 0.1 & 100.0 \\ 
        \bottomrule
    \end{tabular}    
        \vspace{-4ex}
    \label{tab:attack_surface}
\end{table}

\subsection{RQ2. System Efficiency}

We evaluate \toolnospace's impact on end-to-end request execution using the social network microservice benchmark from DeathStarBench~\cite{deathstarbench}. 
We execute a total of 12 million request workflows with 1,000 concurrent requests and complete the experiments within 12 hours. 
To assess robustness across deployment environments, we evaluate three configurations: (i) Docker~\cite{docker} with bridge networking, (ii) a local two-node Kubernetes cluster using Minikube~\cite{minikube} with bridge networking, and (iii) the same Kubernetes setup with Cilium~\cite{cilium} as the container network interface.

\begin{figure*}[t]
    \centering
    \includegraphics[width=.85\textwidth]{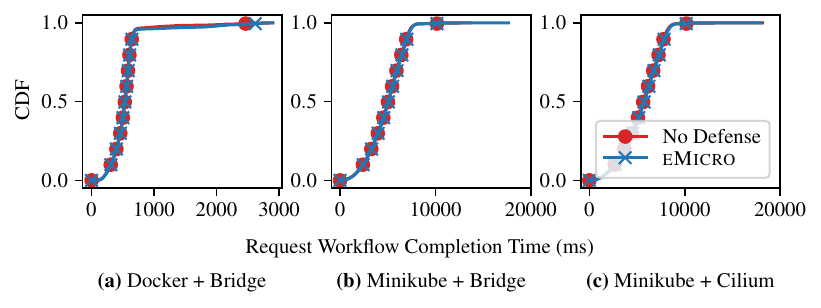}
    \caption{Cumulative distribution function (CDF) for request workflows completion time in various settings. \tool introduces negligible overhead with no significant CDF shift.}
    \label{fig:cdf}
\end{figure*}

\myparagraph{System overhead}
For each request workflow, we record the end-to-end completion time and construct cumulative distribution functions (CDFs) for runs with and without \tool enabled. 
\cref{fig:cdf} shows that \tool does not introduce a significant shift in the latency distribution across orchestration platforms and network configurations.

\myparagraph{Per-component overhead}
To isolate the overhead introduced by individual enforcement components, we measure the additional processing time incurred at each kernel-level hook used by \tool.
These include TC ingress (\verb|__netif_receive_skb|), TC egress (\verb|__dev_queue_xmit|), and three kernel probes to propagate contexts within services (\texttt{inet\_} \texttt{csk\_} \texttt{accept}, \texttt{clone} syscall, \texttt{tcp\_}  \texttt{conn\\ect}). 
\cref{tab:per_component_overhead} shows the baseline processing time of each function and the additional cost introduced by \tool.  
In all cases, the added overhead is on the order of microseconds and remains small relative to the original kernel processing cost.

\begin{figure}[]
    \centering
    \begin{subfigure}[t]{0.48\linewidth}
        \includegraphics[width=\linewidth]{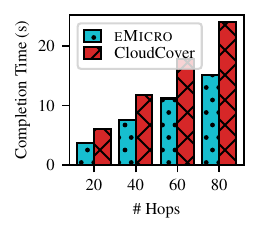}
        \caption{}\label{fig:compare_hops}
    \end{subfigure}
    \hfill
    \begin{subfigure}[t]{0.48\linewidth}
        \includegraphics[width=\linewidth]{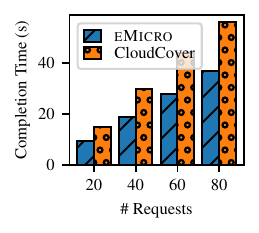}
        \caption{}\label{fig:compare_requests}
    \end{subfigure}
    
    \caption{Comparison of \tool to a major centralized approach across number of \textbf{(a)} hops and \textbf{(b)} requests. \tool requires 0.4s per request and 0.2s per hop, while the centralized approach requires 0.7s and 0.3s, yielding $\sim$35\% improvement.}

\end{figure}

\myparagraph{Comparison to prior approaches}
We compare \tool with CloudCover~\cite{cloudcover}, a representative centralized approach.
We re-implement CloudCover by manually instrumenting the HTTP server library to propagate request context and deploying a centralized service that infers access control decisions.
We measure end-to-end request workflow completion time while varying the number of hops and concurrent requests.
As shown in \cref{fig:compare_hops} and \cref{fig:compare_requests}, \tool consistently outperforms CloudCover by 35\% on average.
In addition, \tool requires only 1~$\mu$s per policy check, compared to 1.6~ms for Kalium~\cite{kalium} and 3~ms for Trapeze.
This advantage comes from fully distributed request tracing and enforcement using eBPF, combined with efficient DFA-based policy encoding.

\begin{table}[]
    \centering
    \small
    \caption{Execution time of each kernel function and the overhead added by attaching the \toolnospace's eBPF probe}
    \begin{tabular}{lcc}
        \toprule
        \textbf{Kernel Function} & \textbf{Time ($ \mu $s)} & \textbf{\tool $\Delta$ ($ \mu $s)}  \\
        \midrule
         \verb|__netif_receive_skb| & 12.6  & 1.4 \\
         \verb|inet_csk_accept| & 4.7 & 2.0 \\
         \verb|__x64_sys_clone| & 291 & 0.7 \\
         \verb|tcp_connect| & 51.7 & 1.9 \\
         \verb|__dev_queue_xmit| & 4.1 & 1.3 \\
         \bottomrule
    \end{tabular}    
    \label{tab:per_component_overhead}
\end{table}

\begin{table}[]
    \centering
    \small
     \caption{Compilation time for \toolnospace's policy language under varying numbers of policies, AST depths, and services}
    \begin{tabular}{cccc}
        \hline
        \textbf{\# Policy} & \textbf{AST Depth} & \textbf{\# Services} & \textbf{Time (s)} \\
        \hline
        \multirow{2}{*}{100} & 2 & \multirow{2}{*}{25} & 0.72 \\
         & 4 & & 1.25 \\
        \hline
        \multirow{2}{*}{1000} & 2 & \multirow{2}{*}{200} & 106.11 \\
        & 4 & & 115.01 \\
        \hline
    \end{tabular}
   
    \label{tab:compiler_eval}
\end{table}

\myparagraph{Compiler efficiency}
We evaluate the efficiency of \toolnospace's policy compiler by measuring the time required to translate high-level policies into a single enforceable DFA. 
For various configurations (number of policies, abstract syntax tree depths, and number of services), we measure the total compilation time, including predicate-to-DFA conversion, union and intersection of composite predicates, and merging all policies into one DFA. Table~\ref{tab:compiler_eval} shows that policy compilation time increases with policy count and AST depth but remains well within practical bounds.

\begin{tcolorbox}[colback=magenta!5, colframe=magenta!80!black,left=3pt,right=3pt,top=3pt,bottom=3pt]
\textbf{Takeaway:} \tool incurs <10~$\mu$s overhead, preserves microservice orchestration performance, compiles 1000 policies in under 2 minutes, and outperforms CloudCover by 35\%.
\end{tcolorbox}

\subsection{RQ3. Label Efficiency}

\begin{table}[t]
    \centering
    \small
    \caption{Production cloud trace datasets. We reconstruct request workflows to measure DFA encoding efficiency.}
    \begin{tabular}{lccc}
        \toprule
        \textbf{Dataset} & \textbf{\# Traces} & \textbf{Duration} & \textbf{Size} \\
        \midrule
        Uber \cite{uber_ds} & 1.4M & 7 days & 570GB \\
        Alibaba \cite{ali_ds,ali_ds2}& 1M & 12 hrs & 49GB \\
        ByteDance \cite{bd_ds}& 1K & N.A.\tablefootnote{The information is not mentioned in the partially released data.}  & 5MB \\
        \bottomrule
    \end{tabular}    
    \label{tab:dataset}
\end{table}

\begin{table}[t]
    \centering
    \small
    \caption{Packet header space required to encode request workflows. \tool reduces this overhead by 90\% relative to the naive encoding baseline.}
    \begin{tabular}{lcccccc}
        \toprule
        \textbf{Dataset} & \textbf{\makecell{\#\\Svc}} & \textbf{\makecell{Max\\hop}} & \textbf{\makecell{\#\\ States}}  & \textbf{\makecell{Naive\\(Bits)}} & \textbf{\makecell{DFA\\(Bits)}} & \textbf{$\Delta$(\%)} \\ 
        \midrule
        SocialNetwork \cite{deathstarbench} & 27 & 5 & 18 & 25 & 5 & -80\% \\
        Media \cite{deathstarbench} & 33 & 6 & 5 & 36 & 4 & -88\% \\
        Uber \cite{uber_ds} & 2.4K & 85 & 260K & 955 & 18 & -98\% \\
        Alibaba \cite{ali_ds,ali_ds2} & 45K & 31 & 1.1M & 480 & 21 & -95\% \\
        ByteDance \cite{bd_ds} & 3.4K & 12 & 27K & 141 & 15 & -89\% \\
        \bottomrule
    \end{tabular}    
    \label{tab:header}
\end{table}

We compare \toolnospace's DFA-based strategy against a naive encoding baseline.
We use cloud trace datasets from Uber \cite{uber_ds}, Alibaba \cite{ali_ds, ali_ds2}, and ByteDance \cite{bd_ds} as shown in \cref{tab:dataset}. 

\myparagraph{Naive encoding} 
To calculate the number of bits required in naive encoding, we analyze the number of unique services and the maximum depth of the request workflow, following \cref{eq:naive}.
\cref{tab:header} shows the results. 
Note that naive encoding can require up to 955 bits (119 bytes) in large-scale microservices, as observed in Uber cloud traces with 2400 services.
This exceeds the standard packet headers such as IP (20-60 bytes), TCP (20-60 bytes), and UDP (8 bytes). Using such encoding would significantly reduce system throughput. 

\myparagraph{DFA encoding}
We generate a DFA from the request workflows reconstructed from each of the cloud trace datasets above and compute the required label size using \cref{eq:dfa}.
\cref{tab:header} shows that the number of states grows exponentially with the number of services.
This is expected since one state represents a distinct request workflow.
However, encoding this state space requires at most $\sim$20~bits, a $\sim$90\% reduction relative to the baseline.

\begin{tcolorbox}[colback=magenta!5, colframe=magenta!80!black,left=3pt,right=3pt,top=3pt,bottom=3pt]
\textbf{Takeaway:} \tool encodes multi-hop policies for 45K services in just 21-bit labels, a 90\% reduction in space overhead.
\end{tcolorbox}

\subsection{RQ4. Scalability of In-Kernel Label Propagation and Policy Enforcement}

To evaluate the scalability of \toolnospace's in-kernel label propagation and policy enforcement, we measure the size of BPF maps and the time taken to build them and perform policy checks, under varying numbers of states and policies.
We use three (\texttt{\#States}, \texttt{Key size (bytes)}, \texttt{Value size (bytes)}) configurations: (256, 2, 1), (65K, 3, 2), and (16M, 4, 3).
Here, the value stores a state identifier using the minimum number of bytes needed to encode all states, the key stores the state plus one additional byte for up to $2^8$ transitions (the smallest size supported by BPF maps that also ensures policy uniqueness for up to 50M policies in our experiments), and the number of policies affects the number of entries.

\begin{figure}[t]
    \centering
    \begin{subfigure}[]{0.5\linewidth}
        \includegraphics[width=\linewidth]{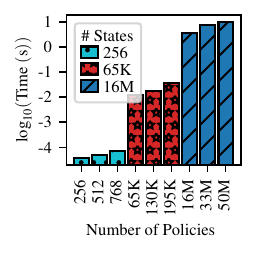}
        \caption{}\label{fig:time_overhead}
    \end{subfigure}
    \hfill
    \begin{subfigure}[]{0.48\linewidth}
        \includegraphics[width=\linewidth]{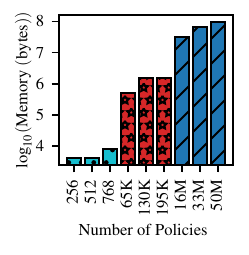}
        \caption{}\label{fig:memory_overhead}
    \end{subfigure}

    \caption{\textbf{(a)} Time and \textbf{(b)} memory overhead of storing multi-hop policies in BPF maps. \tool requires only $<$10 seconds and 100 MB to rebuild and store 50 million policies.}

\end{figure}

\myparagraph{BPF map build and insert time} 
\cref{fig:time_overhead} shows that 50 million multi-hop policies are rebuilt and reinserted into BPF maps in under $\sim$10 seconds, comparable to typical service recovery time such as pulling container images and restarting containers.
This confirms that \tool's recovery cost remains practical at production scale.

\myparagraph{BPF map size} 
\cref{fig:memory_overhead} shows that 50 million policies (16 million DFA states) require only 100~MB of space. 
This efficiency comes from the compact DFA representation of valid request workflows and state minimization: 50 million policies do not translate into 50 million BPF map entries.
The 16-million-state capacity is large enough to cover every public cloud trace we analyzed; the largest, Alibaba~\cite{ali_ds,ali_ds2}, has 45K services and 1.1 million states.
Note that the resulting 100~MB of storage for 50 million entries fits easily in modern memory, without requiring complex mechanisms to split policies across nodes or maintain consistency.

\begin{figure}[t]
    \centering
    \includegraphics[]{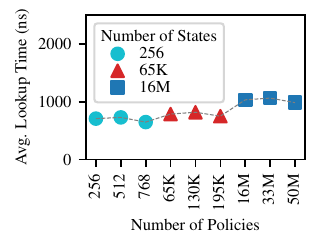}
    \caption{Mean lookup time across varying policy counts. \tool takes only 1~$\mu$s to make an access control decision.}
    \label{fig:lookup}
    \vspace{-1ex}
\end{figure}

\begin{figure}
    \centering
    \begin{subfigure}[]{0.4\linewidth}
        \includegraphics[width=\linewidth]{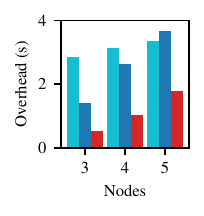}
        \caption{}\label{fig:raft_node}
    \end{subfigure}
    \hfill
    \begin{subfigure}[]{0.55\linewidth}
        \includegraphics[width=\linewidth]{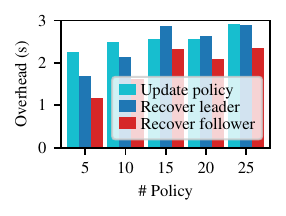}
        \caption{}\label{fig:raft_policy}
    \end{subfigure}
    \caption{Time to propagate policy updates and recover from leader and follower failures across varying cluster sizes and policy counts.}
\end{figure}

\begin{tcolorbox}[colback=magenta!5, colframe=magenta!80!black,left=3pt,right=3pt,top=3pt,bottom=3pt]
\textbf{Takeaway:} \tool distributes policy under 4~s and recovers from node failure in an additional 4~s.
\end{tcolorbox}

\myparagraph{Policy lookup time}
\toolnospace's decentralized design does not require additional network requests for access control decisions.
\cref{fig:lookup} shows that the lookup time remains constant for the same number of states.
Increasing the state count enlarges the BPF map key, slightly raising hash time by a negligible amount ($<$ 200~ns).
Even at 16 million states and 50 million policies, \tool resolves an access control decision in just 1~$\mu$s.
This scalability stems from the DFA's deterministic transition and the constant-time complexity of BPF hash map lookup.

\begin{tcolorbox}[colback=magenta!5, colframe=magenta!80!black,left=3pt,right=3pt,top=3pt,bottom=3pt]
\textbf{Takeaway:} \tool stores 50 million multi-hop policies in 100~MB, rebuilds them in 10~s, and serves constant-time lookups in under 1~$\mu$s.
\end{tcolorbox}

\subsection{RQ5. Distributed Update and Resilience}

We evaluate \toolnospace's ability to support distributed policy updates and recover from node failures using a Raft-replicated policy store.
Policy state is represented as key-value entries identical to those enforced in the kernel: each key is a (current state, next hop) tuple and each value is the corresponding next state, so replicated state applies directly to in-kernel BPF maps without translation.
We deploy a Raft cluster across multiple nodes, elect a leader, and issue policy updates to that leader, measuring the end-to-end latency to replicate and apply each update.
To evaluate fault tolerance, we simulate follower failures by stopping and restarting follower nodes, and leader failures by stopping the leader to trigger re-election.
In both cases, we measure the time to restore policy consistency.

\cref{fig:raft_node,fig:raft_policy} show that policy updates and recovery from both failure types complete within a few seconds across varying cluster sizes and policy counts, demonstrating that distributed enforcement state can be maintained with low operational overhead.


\section{Discussion} \label{sec:discussion}

\myparagraph{Compatibility with connection pooling and multiplexing}
\tool operates in a protocol-agnostic manner.
When an L4 connection carries only one request (i.e., neither pooled nor reused), the connection itself suffices as the request identifier (e.g., a socket file descriptor or 5-tuple).
If connections are pooled or reused, the same identifier can be used but must be refreshed upon reuse.
In deployments that multiplex requests over a single L4 connection (e.g., HTTP/2, HTTP/3), \tool requires a per-request identifier that distinguishes concurrent streams within the connection.
For example, it may be defined as a tuple of the L4 connection identifier and the stream ID (e.g., HTTP/2 stream ID).
More generally, \tool can use any request identifier observable from the eBPF program at the relevant granularity:  socket-level for non-multiplexed traffic, stream-level for multiplexed traffic.

\myparagraph{Compatibility with other networking stacks}
We prototype \tool with eBPF programs attached to TC and XDP because these mechanisms are widely deployed in microservice networks (e.g., Cilium, Calico) and provide low-latency, line-rate processing in the kernel networking stack. 
However, \toolnospace's design is not limited to the kernel networking stack itself. 
In environments that use kernel-bypass networking stacks (e.g., DPDK), the same processing logic (label parsing, DFA transitions, enforcement) can be ported.
For example, packet parsing can be implemented as a DPDK user-space program that extracts and inserts DFA labels. 
It can access the DFA state and transition maps through shared memory (e.g., pinned BPF maps).
Execution unit tracing remains in eBPF programs in the host kernel, since it relies on kernel events (e.g., socket operations, syscalls) that DPDK does not bypass.

\myparagraph{Propagating labels in asynchronous workflows}
\tool naturally supports both synchronous and asynchronous execution because it tracks causality through interactions between execution units and network connections. 
As long as the DFA label is propagated through these channels, \tool preserves it correctly.
When request workflows are mediated by a broker, queue, or shared storage, the causal chain breaks at this boundary.
This is an inherent limitation of all history-based systems, since no observable signal remains to reconstruct the missing dependency. 
In such cases, \tool can be extended with lightweight instrumentation at the broker, queue, or shared storage to attach and propagate labels, preserving causality across these boundaries.

\myparagraph{Policy distribution}
Although our evaluation shows that it is possible for every node to hold the complete 50 million multi-hop policies for large microservice deployments, there may be times when the policies are so large that they must be partitioned across nodes.
A distributed scheme would ensure that each node holds at least the policies for the services assigned to it, while keeping partitions reconstructible in the event of node failure.
Designing such a scheme is beyond the scope of this paper and left for future work.



\section{Conclusion}

\tool is a protocol-agnostic, scalable multi-hop access control system for microservices. It compiles policies into a compact DFA in which each request maps to a state, enabling constant-time checks. This state is propagated by tracing request execution across multi-language microservices using eBPF probes on kernel events and entities. Our evaluations on widely-established benchmarks and production cloud traces show that \tool adds negligible performance overhead and scales to large deployments.
  
We see several promising directions for future work. First, policy checks could be offloaded or performed preemptively in the network using P4-programmable switches, enabling line-rate enforcement and reducing host overhead in multi-node deployments. 
Second, \toolnospace's tracing mechanism can be extended to microservice performance profiling, providing zero-modification instrumentation across heterogeneous languages and protocols. 
Third, \tool can be extended to support more expressive policy models beyond regular languages, such as pushdown automata (PDA), to capture more complex request workflow dependencies.

\section*{Ethical Considerations}

Microservices support critical modern systems, and their misuse has led to publicly reported catastrophic failures. By enforcing multi-hop request workflows rather than only adjacent hops, \tool provides a more secure runtime environment and reduces the likelihood of such failures through fine-grained, end-to-end enforcement using eBPF and compact DFA representations.

Throughout its development, \tool relied exclusively on ethically sourced, publicly available datasets. All experiments were performed in isolated infrastructure under our control, without interacting with or impacting external systems. We recognize that publishing detailed security mechanisms can inform both defenders and adversaries; however, we believe that transparency and open evaluation are essential for scientific integrity and for enabling the community to build safer, more resilient systems

\section*{Open Science}

We provide an implementation of \tool along with documentation to support future research building on our work.
The artifact is available at:
\url{https://github.com/peng-gao-lab/emicro}.

\section*{Acknowledgement}

We would like to thank the anonymous reviewers and our shepherd for their constructive comments.
This work is supported in part by the National Science Foundation under grant 2442171 and the Google Academic Research Award.
Any opinions, findings, and conclusions made in this paper are those of the authors and do not necessarily reflect the views of the funding agencies.
We acknowledge the use of large language models for polishing the manuscript, including grammar, punctuation, and minor edits.
All experimental results, design choices, and the main contributions of the paper remain the sole responsibility of the authors.

\bibliographystyle{ACM-Reference-Format}

\balance
\bibliography{references}

\end{document}